\documentstyle[12pt]{article}

\input epsf.tex

\begin{document}

\author{S. T. R. Pinho, T. A. S. Haddad, and S. R. Salinas \\
Instituto de F\'{\i}sica\\
Universidade de S\~{a}o Paulo\\
Caixa Postal 66318\\
05315-970, S\~{a}o Paulo, SP, Brazil\\
e-mail address: ssalinas@if.usp.br}
\title{Critical behavior of the Ising model on a hierarchical lattice with
aperiodic interactions }
\date{23 September 1997 }
\maketitle

\begin{abstract}
We write exact renormalization-group recursion relations for
near\-est-neighbor ferromagnetic Ising models on Migdal-Kadanoff hierarchical
lattices with a distribution of aperiodic exchange interactions according to
a class of substitutional sequences. For small geometric fluctuations, the
critical behavior is unchanged with respect to the uniform case. For large
fluctuations, as in the case of the Rudin-Shapiro sequence, the uniform
fixed point in the parameter space cannot be reached from any physical
initial conditions. We derive a criterion to check the relevance of the
geometric fluctuations.


\end{abstract}

There are a number of recent investigations on the critical behavior of
ferromagnetic spin systems with the inclusion of aperiodic exchange
interactions \cite{ref1}. In particular, there are detailed studies of the
critical behavior in the ground state of the quantum Ising chain in a
transverse field\cite{ref2,ref3,ref4,ref5} (which is known to be related to
the transition at finite temperatures of the two-dimensional Ising model).
The nearest-neighbor ferromagnetic exchange interactions are chosen
according to some substitution sequences, and the geometric fluctuations are
gauged by a wandering exponent $\omega $ associated with the eigenvalues of
the substitution matrix of the sequence. According to a heuristic criterion
proposed by Luck\cite{ref3}, the critical behavior remains unchanged (that
is, of Onsager type) for bounded fluctuations (small values of $\omega $),
but large fluctuations should induce much weaker singularities, similar to
the case of a disordered Ising ferromagnet.

In a very recent publication\cite{ref6}, we took advantage of the
simplifications brought about by a Migdal-Kadanoff hierarchical (MKH) lattice%
\cite{ref7} to perform some exact calculations for characterizing the
critical behavior of aperiodic ferromagnetic Ising models. The (layered)
exchange interactions between nearest neighbors were chosen according to a
certain class of generalized Fibonacci sequences\cite{ref8}. In this paper,
we review some of these calculations, discuss the well known case of the
Rudin-Shapiro sequence, and present an exact derivation of an analog of
Luck's criterion to check whether the geometric fluctuations are strong
enough to change the critical behavior of the uniform system.

Consider a particular two-letter generalized Fibonacci sequence given by the
substitutions 
\begin{equation}
\begin{array}{l}
A\rightarrow AB, \\ 
B\rightarrow AA.
\end{array}
\label{eq1}
\end{equation}
If we start with letter $A$, the successive application of this inflation
rule produces the sequences 
\begin{equation}
A\rightarrow AB\rightarrow ABAA\rightarrow ABAAABAB\rightarrow \cdot \cdot
\cdot .  \label{eq2}
\end{equation}
At each stage of this construction, the numbers $N_A$ and $N_B$, of letters $%
A$ and $B$, can be obtained from the recursion relations 
\begin{equation}
\left( 
\begin{array}{c}
N_A^{\prime } \\ 
N_B^{\prime }
\end{array}
\right) ={\bf M}\left( 
\begin{array}{c}
N_A \\ 
N_B
\end{array}
\right) ,  \label{eq3}
\end{equation}
with the substitution matrix 
\begin{equation}
{\bf M}=\left( 
\begin{array}{cc}
1 & 2 \\ 
1 & 0
\end{array}
\right) .  \label{eq4}
\end{equation}
The eigenvalues of this matrix, $\lambda _1=2$ and $\lambda _2=-1$, govern
most of the geometrical properties. For any sequence, the total number of
letters, at a large order $n$ of the construction, depends asymptotically on 
$\lambda _1^n$. The fluctuations are of order $\left| \lambda _2\right| ^n$.
Then, it is interesting to define the wandering exponent\cite{ref3}, 
\begin{equation}
\omega =\frac{\ln \left| \lambda _2\right| }{\ln \lambda _1},  \label{eq8}
\end{equation}
that expresses the asymptotic dependence of the fluctuations with the total
number of letters, $\Delta N^{\left( n\right) }\sim N^\omega $.

The nearest-neighbor Ising model is given by the Hamiltonian 
\begin{equation}
{\cal H}=-\sum\limits_{\left( i,j\right) }J_{i,j}\sigma _i\sigma _j,
\label{eq9}
\end{equation}
with the spin variables $\sigma _i=\pm 1$ on the sites of a hierarchical
diamond structure. In Fig. 1, which is suitable for the period-doubling
Fibonacci rule of Eq.(\ref{eq1}), we draw the first stages of the
construction of a diamond lattice with a basic polygon of four bonds (that
is, of a MKH lattice with cell length $b=2$, and number of branches $q=2$,
which amounts to $4$ bonds in the diamond unit cell). As indicated in this
figure, we simulate a layered system by the introduction of the interactions 
$J_A>0$ and $J_B>0$ along the branches of the structure. From the rules of
Eq. (\ref{eq1}), it is straightforward to establish the recursion relations 
\begin{equation}
x_A^{\prime }=\frac{2x_Ax_B}{1+x_A^2x_B^2},  \label{eq12}
\end{equation}
and 
\begin{equation}
x_B^{\prime }=\frac{2x_A^2}{1+x_A^4},  \label{eq13}
\end{equation}
where $x_A=\tanh K_A$, $x_B=\tanh K_B$, $K_A=\beta J_A$, $K_B=\beta J_B$,
and $\beta $ is the inverse of the temperature.

It should be remarked that similar procedures can be used to consider much
more general substitutional sequences. However, to avoid any changes in the
topology of the hierarchical lattice, we restrict the analysis to
period-multiplying substitutions. In these cases, the largest eigenvalue of
the inflation matrix, $\lambda _1$, gives the multiplication factor of the
period. Therefore, $b=\lambda _1$ for all cases under consideration (and, in
particular, $b=\lambda _1=2$, for the diamond lattice of the figure).

\[ \epsfbox{fig1.eps} \]
\begin{footnotesize}
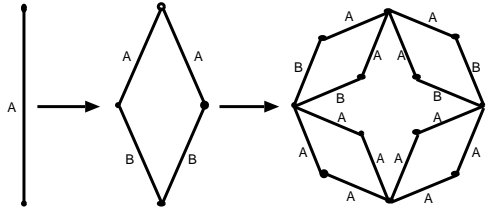
{\bf Fig. 1-} Some stages of the construction of a Midgal-Kadanoff
hierarchical lattice with bond length $b=2$ and $q=2$ branches (which
is called a diamond lattice) for the period-doubling sequence $A
\rightarrow AB$ and $B \rightarrow AA$ (letters $A$ and $B$ indicate
the exchange interactions, $J_A$ and $J_B$).  
\end{footnotesize}
\vspace{0.57cm}

In the uniform case, $J_A=J_B=J$, Eqs. (\ref{eq12}) and (\ref{eq13}) reduce
to the simple recursion relation 
\begin{equation}  \label{eq14}
x^{\prime }=\frac{2x^2}{1+x^4},
\end{equation}
with two trivial and stable fixed points, $x^{*}=0$ and $x^{*}=1$, and a
non-trivial and unstable fixed point, $x^{*}=0.543689...$, which come from
the polynomial equation 
\begin{equation}  \label{eq15}
x^5-2x^2+x=x\left[ x^4-2x+1\right] =0.
\end{equation}

In the aperiodic case ($J_A\neq J_B$) under consideration, the $A$
components of the coordinates of the fixed points in the physical sectors of
the $x_A-x_B$ space ($0\leq x_A,x_B\leq 1$) come from the solutions of the
equation 
\begin{equation}
x_A^9+2x_A^5-4x_A^3+x_A=x_A\left[ x_A^4-2x_A+1\right] \left[
x_A^4+2x_A+1\right] =0.  \label{eq16}
\end{equation}
Therefore, a comparison with Eq. (\ref{eq15}) shows that the only fixed
points are located along the $x_A=x_B$ direction, and given by the same
values as in the uniform case, $x_A^{*}=x_B^{*}=0$, $x_A^{*}=x_B^{*}=1$, and 
$x_A^{*}=x_B^{*}=0.543689...$. The linearization about the non-trivial
uniform fixed point yields the asymptotic equations 
\begin{equation}
\left( 
\begin{array}{l}
\Delta x_A^{\prime } \\ 
\Delta x_B^{\prime }
\end{array}
\right) =C{\bf M}^T\left( 
\begin{array}{l}
\Delta x_A \\ 
\Delta x_B
\end{array}
\right) ,  \label{eq17}
\end{equation}
where ${\bf M}^T$ is the transpose of the substitution matrix, and the
structure factor $C$ is given by 
\begin{equation}
C=\frac{1-x_A^{*}}{x_A^{*}}=0.839286....  \label{eq18}
\end{equation}
The diagonalization of this linear form gives the eigenvalues 
\begin{equation}
\Lambda _1=C\lambda _1=2C=1.678573...,  \label{eq19}
\end{equation}
and 
\begin{equation}
\Lambda _2=C\lambda _2=-C=-0.839286....,  \label{eq20}
\end{equation}
where $\lambda _1=2$ and $\lambda _2=-1$ are the eigenvalues of the
substitution matrix. As $\Lambda _1>1$ and $\left| \Lambda _2\right| <1$,
the fixed point is a saddle node with a flipping approximation. Therefore,
given the ratio $r=J_B/J_A$ between the exchange interactions, the critical
temperature is defined by the flow into this uniform fixed point. From Eqs. (%
\ref{eq14}) and (\ref{eq15}), we see that the same eigenvalue $\Lambda _1$
characterizes the (unstable) fixed point of the uniform model. Thus, in this
particular example, with the wandering exponent $\omega =0$, the geometric
fluctuations are unable to change the critical behavior with respect to the
uniform system. We can draw a phase diagram, in terms of the ratio $r$ and
the temperature $T$, where the critical line displays the same (universal)
exponents of the uniform case. Also, it is not difficult to check that the
same sort of behavior (saddle point; largest eigenvalue associated with the
uniform system) still holds for all finite values of the branching number $q$
of the diamond ($b=2$) structure.

To give an example with another value of the wandering exponent $\omega $,
where the geometric fluctuations become relevant, consider the (four-letter)
Rudin-Shapiro sequence\cite{ref8}, $A\rightarrow AC$, $B\rightarrow DC$, $%
C\rightarrow AB$, and $D\rightarrow DB$. The substitution matrix is given by 
\begin{equation}
{\bf M}_{RS}=\left( 
\begin{array}{cccc}
1 & 0 & 1 & 0 \\ 
0 & 0 & 1 & 1 \\ 
1 & 1 & 0 & 0 \\ 
0 & 1 & 0 & 1
\end{array}
\right) ,  \label{eq21}
\end{equation}
with eigenvalues $\lambda _1=2$, $\lambda _2=-\lambda _3=\sqrt{2}$, and $%
\lambda _4=0$, and the wandering exponent $\omega =1/2$. For the Ising model
on the diamond hierarchical lattice of the figure, it is easy to write the
set of recursion relations 
\begin{equation}
x_A^{\prime }=\frac{2x_Ax_C}{1+x_A^2x_C^2};\medskip\ x_B^{\prime }=\frac{%
2x_Dx_C}{1+x_D^2x_C^2};  \label{eq22}
\end{equation}
\begin{equation}
x_C^{\prime }=\frac{2x_Ax_B}{1+x_A^2x_B^2};\medskip\ x_D^{\prime }=\frac{%
2x_Dx_B}{1+x_D^2x_B^2}.  \label{eq23}
\end{equation}
Again, there are two trivial fixed points, $%
x_A^{*}=x_B^{*}=x_C^{*}=x_D^{*}=0 $ and $1$, and the non-trivial uniform
fixed point, $x_A^{*}=x_B^{*}=x_C^{*}=x_D^{*}=0.543689...$, as in the
uniform case. The linearization about this uniform fixed point gives a
matrix relation of the same form as Eq. (\ref{eq17}), 
\begin{equation}
\left( {\bf \Delta x^{\prime }}\right) =C{\bf M}_{RS}^T\left( {\bf \Delta x}%
\right) ,  \label{eq24}
\end{equation}
with the same structure factor $C$, given by Eq. (\ref{eq18}), and the
eigenvalues 
\begin{equation}
\Lambda _1=C\lambda _1=2C=1.678573...,  \label{eq25}
\end{equation}
\begin{equation}
\Lambda _{2,3}=C\lambda _{2,3}=\pm C\sqrt{2}=\pm 1.186930...,  \label{eq26}
\end{equation}
and 
\begin{equation}
\Lambda _4=C\lambda _4=0.  \label{eq27}
\end{equation}
Therefore, besides being unstable along the diagonal direction ($%
x_A=x_B=x_C=x_D$), this uniform fixed point is also unstable along two
additional directions in the four-dimensional $x_A-x_B-x_C-x_D$ parameter
space. Given the ratios between the exchange interactions, there is no
temperature associated with any physical initial conditions in this
parameter space so that we can reach the uniform fixed point. The critical
behavior is of a much more subtle character as compared with the uniform
case.

Now it is interesting to devise an analog of Luck's criterion to gauge the
influence of the geometric fluctuations. As we have seen in the previous
examples, the largest eigenvalue in the neighborhood of the uniform fixed
point is given by 
\begin{equation}
\Lambda _1=\lambda _1C=bC,  \label{eq28}
\end{equation}
where it is important to remark that the calculations are always performed
for substitutional sequences and MKH lattices such that $\lambda _1=b$. The
second largest eigenvalue is given by 
\begin{equation}
\Lambda _2=\lambda _2C=\frac{\lambda _2}b\Lambda _1.  \label{eq29}
\end{equation}
Therefore, the fluctuations are relevant if 
\begin{equation}
\left| \Lambda _2\right| =\frac{\left| \lambda _2\right| }b\Lambda _1>1.
\label{eq30}
\end{equation}
>From the exact recursion relations between the free energies associated with
successive generations of a uniform ferromagnetic Ising model on a MKH
lattice\cite{ref7}, we can write 
\begin{equation}
\Lambda _1=b^{y_t}=b^{\frac D{2-\alpha }},  \label{eq31}
\end{equation}
where $\alpha $ is the critical exponent of the specific heat of the uniform
model and $D$ is the fractal dimension of the lattice. From the definition
of the wandering exponent, given by Eq. (\ref{eq8}), we can also write 
\begin{equation}
\left| \lambda _2\right| =b^\omega .  \label{eq32}
\end{equation}
Inserting these expressions into inequality (\ref{eq30}), we show that the
geometric fluctuations are relevant for 
\begin{equation}
\omega >\omega _c=1-\frac D{2-\alpha }.  \label{eq33}
\end{equation}

In the particular case of the diamond hierarchical lattice ($b=2$ and $q=2$)
the fractal dimension is given by 
\begin{equation}
D=\frac{\ln \left( qb\right) }{\ln b}=2,  \label{eq34}
\end{equation}
so the criterion is reduced to the inequality 
\begin{equation}
\omega >\omega _c=-\frac \alpha {2-\alpha }.  \label{eq35}
\end{equation}
As $\Lambda _1=2C=1.678573...$, we have $\alpha =-0.676533...$, and $\omega
>0.252764...$, which explains the universal behavior of the first example ($%
\omega =0$) and the relevance of the fluctuations in the case of the
Rudin-Shapiro sequence ($\omega =1/2$). This same criterion explains the
change in the critical behavior of an aperiodic Potts model on the diamond
hierarchical lattice above $4+2\sqrt{2}$ states, as recently shown by
Magalh\~{a}es, Salinas, and Tsallis\cite{ref9}.

{\bf Acknowledgments}

We thank discussions with R. F. S. Andrade and E. M. F. Curado. STRP is on
leave from Instituto de F\'{\i}sica, Universidade Federal da Bahia, and
thanks a fellowship from the program CAPES/PICD. This work has been
supported by grants from the Brazilian organizations Fapesp and CNPq.

\newpage\ 

\begin{center}
{\bf Figure Caption}
\end{center}

\bigskip\ 

{\bf Fig. 1-} Some stages of the construction of a Migdal-Kadanoff
hierarchical lattice with bond length $b=2$ and $q=2$ branches (which is
called a diamond lattice) for the period-doubling sequence $A\rightarrow AB$
and $B\rightarrow AA$ (letters $A$ and $B$ indicate the exchange
interactions, $J_A$ and $J_B$).

\end{document}